%% file: 00_main.tex
\documentclass{article}

\input{01_preambule_preprint}

\input{02_title_preprint}

\begin{document}

\maketitle

\blfootnote{This work was supported by the MRC and Spinal Research Charity through the ERA-NET Neuron joint call (MR/R000050/1). The Wellcome Centre for Human Neuroimaging is supported by core funding from the Wellcome [203147/Z/16/Z].}

\input{03_abstract}


\input{04_introduction}
\input{05_methods}
\input{06_results}
\input{07_discussion}
\input{08_conclusions}
\bibliography{00_main}
\input{10_appendix}


\end{document}

%% file: 01_preambule_preprint.tex
\usepackage{authblk}
\usepackage{fullpage}
\usepackage{balance}
\usepackage[numbers]{natbib}
\newcommand\blfootnote[1]{%
  \begingroup
  \renewcommand\thefootnote{}\footnote{#1}%
  \addtocounter{footnote}{-1}%
  \endgroup
}

\providecommand{\keywords}[1]
{
  \small	
  \textbf{\textit{Keywords---}} #1
}

\newcommand*{\revised}[1]{#1}                  

\usepackage{graphicx} 
\graphicspath{{./figures/}}

\usepackage{siunitx}    
\usepackage{multicol}   
\usepackage{caption}    
\usepackage{widetable}      
\usepackage{booktabs}       
\usepackage[colorlinks]{hyperref}
\usepackage{float}

\usepackage{amsmath}    
\usepackage{amssymb}    
\usepackage{nicefrac}   
\usepackage{stmaryrd}   

\usepackage{cleveref}   

\newcommand*{\T}{^\mathrm{T}}

\newcommand*{\cequal}{\overset{c}{=}}

\renewcommand*\vec[1]{\mathbf{\boldsymbol{#1}}}


%% file: 02_title_preprint.tex
\title{Correcting inter-scan motion artefacts \\in quantitative $R_1$ mapping at 7T.}

\renewcommand\Authfont{\scshape\normalsize}
\renewcommand\Affilfont{\itshape\small}

\author[1,2]{Ya\"el~Balbastre}
\author[1,3]{Ali~Aghaeifar}
\author[1,4]{Nad\`ege~Corbin}
\author[1]{Mikael~Brudfors}
\author[1]{John~Ashburner}
\author[1]{Martina~F.~Callaghan\footnote{\textbf{Corresponding author: \url{m.callaghan@ucl.ac.uk}}}}

\affil[1]{Wellcome Centre for Human Neuroimaging, UCL Queen Square Institute of Neurology, \protect\\
University College London, London, UK}
\affil[2]{Athinoula A. Martinos Center for Biomedical Imaging, \protect\\
Massachusetts General Hospital and Harvard Medical School, Boston, USA}
\affil[3]{MR Research Collaborations, Siemens Healthcare Limited, Frimley, UK}
\affil[4]{Centre de Résonance Magnétique des Systèmes Biologiques, UMR5536, \protect\\
CNRS/University Bordeaux, Bordeaux, France}

\date{}

%% file: 03_abstract.tex
\begin{abstract}
\textbf{Purpose:} Inter-scan motion is a substantial source of error in $R_1$ estimation methods \revised{based on multiple volumes, e.g. variable flip angle (VFA)}, and can be expected to increase at 7T where $B_1$ fields are more inhomogeneous. The established correction scheme does not translate to 7T since it requires a body coil reference. Here we introduce two alternatives that outperform the established method.  Since they compute relative sensitivities they do not require body coil images.  \\
\textbf{Theory:} The proposed methods use coil-combined magnitude images to obtain the relative coil sensitivities. The first method efficiently computes the relative sensitivities via a simple ratio; the second by fitting a more sophisticated generative model.\\
\textbf{Methods:} $R_1$ maps were computed using the VFA approach. Multiple datasets were acquired at 3T and 7T, with and without motion between the acquisition of the VFA volumes. $R_1$ maps were constructed without correction, with the proposed corrections, and (at 3T) with the previously established correction scheme. \revised{The effect of the greater inhomogeneity in the transmit field at 7T was also explored by acquiring $B_1^+$ maps at each position.}\\
\textbf{Results:} At 3T, the proposed methods outperform the baseline method. Inter-scan motion artefacts were also reduced at 7T. However, reproducibility only converged on that of the no motion condition if position-specific transmit field effects were also incorporated.\\
\textbf{Conclusion:} The proposed methods simplify inter-scan motion correction of $R_1$ maps and are applicable at both 3T and 7T, where a body coil is typically not available. The open-source code for all methods is made publicly available.


\keywords{qMRI, inter-scan motion, sensitivity, generative modelling, $R_1$, 7T}

\end{abstract}

%% file: 04_introduction.tex
    \section{Introduction}

Quantitative MRI, and the push towards \emph{in vivo} histology, aims to extract tissue-specific parameters from a series of weighted volumes \citep{weiskopf2021quantitative}. For example, the longitudinal relaxation rate, $R_1$, which is sensitive to important biological features, such as myelin and iron content, can be quantified with the variable flip angle (VFA) approach, e.g. \citet{deoni2003rapid, helms2008quantitative}. A common assumption when computing quantitative metrics is that certain multiplicative factors, such as the signal intensity modulation imposed by the receiver coil's net sensitivity profile, are constant across the weighted volumes. However, this is invalid if motion occurs between the volume acquisitions. In the case of neuroimaging, rigid body co-registration can be used to realign the brain but will not correct for the differential coil sensitivity modulation, which in $R_1$ maps computed with the VFA approach can lead to mean absolute error approaching 20\% \citep{papp2016correction}.

A correction scheme has previously been proposed by \citet{papp2016correction} and validated for $R_1$ mapping at 3T. The position-specific net receive sensitivity is estimated from two rapid low-resolution magnitude images, received on the body and array coils respectively prior to each VFA acquisition. The more homogeneous profile of the body coil is used as a reference to compute the net receiver sensitivity, which is then removed from the VFA acquisitions. This approach effectively assumes that the body coil's modulation is consistent across volumes instead of that of the array coil. This in itself is a potential limitation, as is the general unavailability at body coils at higher field strengths.  

Here we propose an alternative whereby we estimate the \emph{relative} sensitivity \emph{between} volumes. This approach does not fully remove the receiver's sensitivity modulation but does remove the bias that differential modulation introduces in quantitative metrics. Only the calibration images obtained with the array coil are required, i.e. less data than the originally proposed method \citep{papp2016correction}. To validate the approach, we focus on $R_1$ maps computed with the multi-parameter mapping (MPM) protocol \citep{weiskopf2013quantitative}. We first compare performance with the established method of Papp \emph{et al.} at 3T \citep{papp2016correction} and then demonstrate a reduction of inter-scan motion artefacts at 7T under a range of different motion conditions. We further demonstrate that, unlike at 3T, the transmit field $B_1^+$ also exhibits substantial position-specific variability at 7T. As a result, the most precise $R_1$ estimates were obtained by accounting for both position-specific \revised{transmit and receive sensitivity} effects.

\revised{While we validate this approach in the context of $R_1$ mapping, it has much more general potential and can be applied to other mapping methods that combine data from multiple volumes.}


%% file: 05_methods.tex
\section{Methods}

\subsection{Theory}

$R_1$ mapping can be achieved by acquiring spoiled gradient echo volumes with at least two different flip angles \citep{deoni2003rapid,helms2008quantitative}. At a given spatial location, the image intensity, $I$, for a given nominal flip angle $\alpha$ is:
\begin{equation}
    I_k =  s_k \rho \frac{1 - \exp(-\text{TR}_k R_1)}{1 - \cos(f_{\mathrm{T}_k}\alpha_k) \exp(-\text{TR}_k R_1)} ~,
\end{equation}
where $s$ is the receive sensitivity, $\rho$ is the proton density, $f_\mathrm{T}$ is the transmit field, $R_1$ is the longitudinal relaxation rate, TR is the repetition time, and $k$ indexes the VFA acquisition. Co-registration allows for inter-scan motion by realigning anatomical structure across acquisitions. Under the small flip angle approximation \citep{helms2008quantitative}, with two nominal flip angles ($k$ = \{1,2\}),  $R_1$ can be computed as follows:
\begin{align}
    R_1 = \frac{1}{2}\frac{\frac{s_2I_2f_{\mathrm{T}_2}\alpha_{2}}{\text{TR}_2} - \frac{s_1I_1f_{\mathrm{T}_1}\alpha_{1}}{\text{TR}_{1}}}{\frac{s_1I_1}{f_{\mathrm{T}_1}\alpha_{1}} - \frac{s_2I_2}{f_{\mathrm{T}_2}\alpha_{2}}} ~.
    \label{eq:r1smallfa}
\end{align}
Typically, it is assumed that $s_1 = s_2$ and the sensitivities simplify out. However, this assumption is invalid if inter-scan motion has occurred leading to substantial bias in $R_1$ \revised{estimates} \citep{papp2016correction}.  This can be avoided by accounting for the \emph{relative} sensitivity across positions: $\Delta_{1,2} = s_1 / s_2$. Substitution for $s_1$ in equation \eqref{eq:r1smallfa} gives:
\begin{align}
    R_1 = \frac{1}{2}\frac{\frac{I_2f_{\mathrm{T}_2}\alpha_{2}}{\text{TR}_2} - \frac{\Delta_{1,2}I_1f_{\mathrm{T}_1}\alpha_{1}}{\text{TR}_{1}}}{\frac{\Delta_{1,2}I_1}{f_{\mathrm{T}_1}\alpha_{1}} - \frac{I_2}{f_{\mathrm{T}_2}\alpha_{2}}} ~,
    \label{eq:r1relative}
\end{align}
The method of \citet{papp2016correction} did not include the relative sensitivity but referenced to an additional calibration image acquired on the body coil, assuming that the body coil modulation was position-independent.

It is commonly assumed that the transmit field is sufficiently smooth as to be considered position-independent, \emph{i.e.} $f_{\mathrm{T}_1} = f_{\mathrm{T}_2}$, such that:
\begin{align}
    R_1 = \frac{f_\mathrm{T}^2}{2}\frac{\frac{I_2\alpha_{2}}{\text{TR}_2} - \frac{\Delta_{1,2}I_1\alpha_{1}}{\text{TR}_{1}}}{\frac{\Delta_{1,2}I_1}{\alpha_{1}} - \frac{I_2}{\alpha_{2}}} ~,
    \label{eq:r1singleb1}
\end{align}
However, in this work we show that this assumption does not hold at 7T, and that incorporating position-specific transmit field estimates maximises the precision of $R_1$.


\begin{figure}
    \centering
    \includegraphics[width=0.7\textwidth]{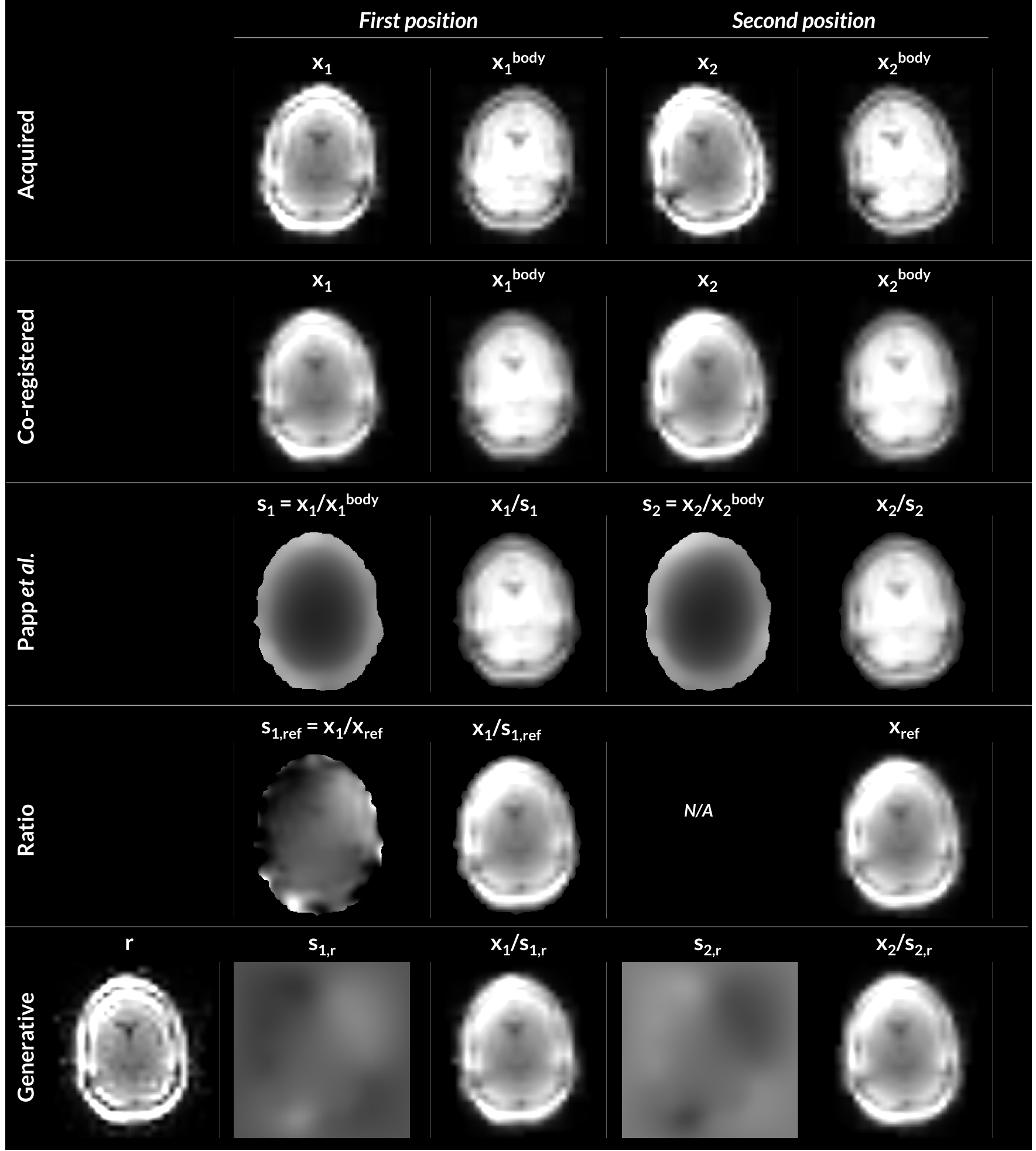}
    \caption{\revised{3T Example. The acquired calibration images, $x_k$ have different orientation due to participant movement between acquisitions (1st row).  Co-registration can align the images spatially, but does not correct for their differential sensitivity field modulation (2nd row), visible via their ratio, $x_1/x_2 = \Delta_{1,2}$. The method from \citet{papp2016correction} estimates and corrects this modulation using an additional body-coil image (3rd row). When one is not available, relative signal differences can be corrected for using the relative modulation $\Delta_{1,2}$ (4th row). Alternatively, the joint log-likelihood of a generative forward model that embeds the spatial transformation from a mean image, $r$, to native space can be maximised to determine the mean image and modulating sensitivities, $s_{k,r}$, that best explain the acquired images $x_k$ (5th row). The generative modelling approach produces a similar relative modulation ($s_{1,r}/s_{2,r} = \Delta_{1,2}$) but allows for the corrected images to have the minimal modulation of the mean image.}}
    \label{fig:sens}
\end{figure}

\textbf{Ratio approach} The calibration data used to correct inter-scan motion artefacts comprised rapid low resolution, coil-combined magnitude images acquired immediately prior to each high resolution VFA acquisition. These images, $\left\{x_k\right\}_{k=1}^K$, assumed to have been rigidly co-registered \revised{to the same space}, can be written as the product of a common image $r$ and a net sensitivity field  $\left\{s_k\right\}_{k=1}^K$. The relative sensitivity, \revised{$\kappa_{k,\text{ref}}$} can be computed with respect to one of the calibration acquisitions, used as a reference:
\revised{\begin{equation}
    \kappa_{k,\text{ref}} = \frac{x_{k}}{x_{\text{ref}}} =\frac{s_k r}{s_{\text{ref}} r} =  \frac{s_k}{s_{\text{ref}}} = \Delta_{k,\text{ref}}
    \label{eq:relative}
\end{equation}}


Dividing each VFA acquisition by its relative sensitivity $\Delta_{k,\text{ref}}$ results in a common modulation, $s_{\text{ref}}$, which, although less homogeneous than the body coil used by Papp \emph{et al.}, more faithfully restores the validity of assuming common modulation when computing $R_1$. 

\revised{The assumption that $r$ is common, such that $\kappa_{k,\text{ref}}$ = $\Delta_{k,\text{ref}}$ holds only if there are no position-specific transmit field effects. Simulations were used to explore the validity of this assumption.}

\textbf{Generative approach} This ratio approach risks noise amplification, particularly in regions of low signal-to-noise ratio (SNR). This is combatted by isotropically smoothing $x_k$ and $x_{\text{ref}}$ before taking their ratio. A potentially more robust alternative is to cast the computation of the relative coil sensitivities, and a common image modulated by them, as an inference problem in a probabilistic generative model of $x_k$ that incorporates noise and can also embed knowledge about the spatial smoothness of the sensitivities. \revised{This generative modelling approach allows coils with arbitrary sensitivity to be incorporated, e.g. coils with more (array) or less (body) spatial variation, or both concurrently ("array + body") if available. \emph{A priori} knowledge about the expected smoothness of the sensitivity can be incorporated at the level of coil type (body versus array) via appropriate tailoring of a regularisation parameter, $\lambda$. Images acquired with the body coil will have a flatter sensitivity field modulation, which can be incorporated by setting $\lambda_{\text{body}} \gg \lambda_{\text{array}}$.} Full details are given in Appendix A.

\subsection{Experiment}

\textbf{Participants} One participant (F, 31 years) was scanned at 3T (MAGNETOM Prisma, Siemens, Erlangen, Germany) using a body coil for transmission and either the body coil or a 32-channel head array coil for reception. Three additional participants (2F, 1M; 32 - 41 years) were scanned at 7T (MAGNETOM Terra, Siemens, Erlangen, Germany) using an 8-channel transmit, 32-channel receive head array coil (Nova Medical, Wilmington, MA, USA) in \revised{a quadrature-like ("TrueForm")} mode. All data were acquired with approval from the UCL research ethics committee.

\textbf{MPM Datasets} MPM data were acquired using a multi-echo spoiled gradient echo sequence with flip angles of 6$^\circ$ (PD-weighted, "PDw") and 26$^\circ$ (T1-weighted, "T1w"), a TR of 19.5 ms and an RF spoiling increment of 117$^\circ$ with a total dephasing gradient moment per TR of 6$\pi$. Eight echoes were acquired with TE ranging from 2.56 ms to 15.02 ms in steps of 1.78 ms using a bandwidth of 651 Hz/pixel. Data were acquired with \revised{a nominal} 1 mm isotropic resolution over a field of view of 160 mm right-left and 192 mm in the anterior-posterior and superior-inferior directions. Elliptical sampling and partial Fourier, with factor 6/8 in each phase-encoded direction, were used to accelerate the acquisition, leading to a scan time of 5 minutes per volume. A $B_1^+$ map was estimated by acquiring a series of spin and stimulated echoes using previously described 3T and 7T protocols \citep{lutti2010optimization,lutti2012robust}. \revised{These data were acquired with 4 mm isotropic resolution resulting in a total acquisition time of 3 minutes 48 seconds, and a further 1 minute for $B_0$ mapping.}

For inter-scan motion correction, additional single echo acquisitions were acquired prior to each VFA acquisition with a flip angle of 6$^\circ$, TE = 2.4 ms, TR = 6.5 ms, a bandwidth of 488 Hz/pixel and no acceleration schemes. At 3T, these data were acquired, receiving sequentially on the array and body coils, with 8 mm isotropic resolution leading to a scan time of 6 s per volume. To capture the greater spatial variation in the net sensitivity field at 7T, the resolution was increased to 4 mm isotropic leading to a scan times of 18 s per volume, but acquired only on the array coil due to the absence of a body coil. 

\textbf{Motion Conditions} Two MPM datasets were acquired to define baseline reproducibility. Participants were then instructed to move to a new, arbitrary position within the confines of the coil.  A localiser was acquired and the field of view repositioned as necessary to ensure appropriate brain coverage in the new position.  A third MPM data set was then acquired. 

\textbf{$B_1^+$ per contrast} \revised{For each $R_1$ map computed from data across two positions, \emph{i.e.} with inter-scan motion, two different corrections for transmit field inhomogeneity were performed. The first assumed the transmit field was identical across head positions (Eq. \eqref{eq:r1singleb1}) and in the same position as the PD-weighted volume. The second used position-specific $B_1^+$ maps (Eq. \eqref{eq:r1relative})}.

\textbf{$R_1$ Analysis} $R_1$ maps were computed using the hMRI toolbox \citep{tabelow2019hmri}, which uses the small flip angle approximation \citep{helms2008quantitative} and corrects for imperfect spoiling \citep{preibisch2009influence}, which always used the $B_1^+$ map acquired in the space of the PDw acquisition.  Maps were computed with and without inter-scan motion using all possible PDw and T1w combinations. To ease comparisons, all maps were constructed in the space of the first PDw volume. Rigid transformations between all volumes and the first PDw volume were estimated using SPM12 (\href{https://www.fil.ion.ucl.ac.uk/spm/}{Wellcome Centre for Human Neuroimaging}) having first corrected for intensity non-uniformity and skull-stripped the images. $R_1$ maps were computed with and without the proposed inter-scan motion correction schemes. At 3T, $R_1$ maps were also computed with the method of \citet{papp2016correction}. An isotropic kernel of 12mm full-width-at-half-maximum -- the default in the hMRI toolbox -- was used to smooth the calibration images prior to computing the relative (proposed) and absolute (Papp \emph{et al.}) sensitivities. The generative modelling approaches used \revised{$\lambda_{\text{array}} = 10^7$, $\lambda_{\text{body}} = 10^9$}, and 15 iterations. 

\textbf{Error metric} \revised{Three (two from position one and the third from position two) $R_1$ maps, with no additional inter-scan motion corrections applied, were averaged to produce a `ground-truth' map, $\hat{R}_1$. All of the available $R_1$ maps for each participant and condition (motion/no motion) were assessed against this reference to quantify the error, and its variability.  The set of $R_1$ maps used to compute the reference were} also segmented to create a mask selecting those voxels with a mean probability of being in WM, GM or CSF greater than 50\%. For participant 3, the cerebellum was excluded, using the SUIT toolbox \citep{DIEDRICHSEN2006127} in SPM, as a result of $B_1^+$ mapping failure caused by excessively large off-resonance. For the $N$ voxels within the resulting participant-specific mask, the mean absolute error, MAE, for each $R_1$ map was computed with respect to the `ground-truth' map, $\hat{R}_1$ as:
\begin{equation}
    \text{MAE} = \frac{1}{N}\sum_{n=1}^N \frac{\left|\hat{R}_1(n) - R_1(n)\right|}{\hat{R}_1(n)} ~.
\end{equation}
These errors are reported as percentages. 

\revised{\subsection{Simulation study}}

\begin{figure*}
    \centering
    \includegraphics[width=0.5\textwidth]{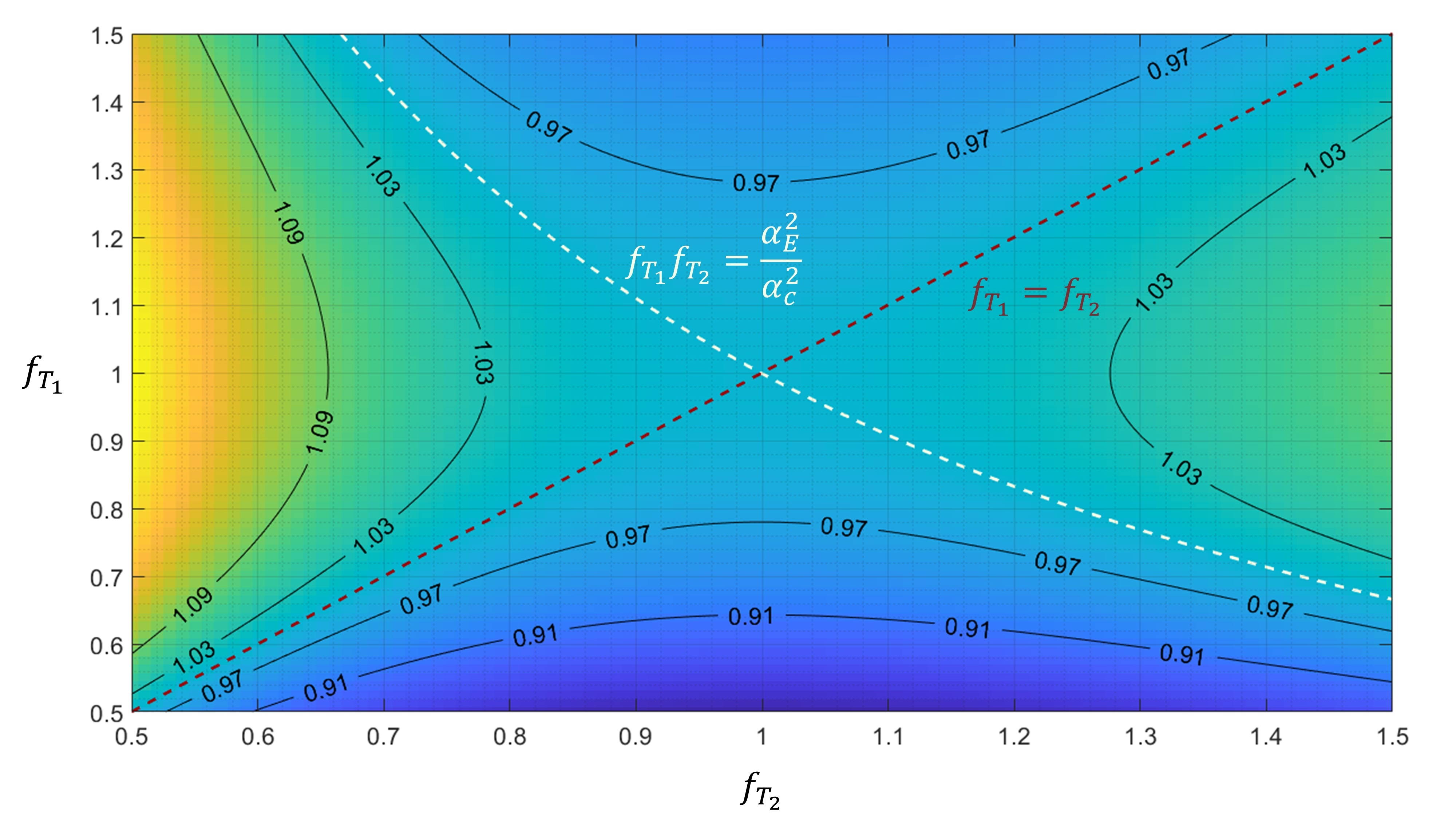}
    \caption{\revised{Misestimation of the true relative sensitivity ($\Delta_{1,2}$) by the ratio of calibration images ($\kappa_{1,2}$), as a function of transmit fields. Colours encode $\frac{\partial \kappa}{\partial\Delta_{1,2}}$, which is 1 when either $f_{\mathrm{T}_1} = f_{\mathrm{T}_2}$ (black doted line) or $f_{\mathrm{T}_1}f_{\mathrm{T}_2}=\alpha_E^2/\alpha_c^2$ (white dotted line).}}
    \label{fig:kappa_sens}
\end{figure*}

\revised{\textbf{Validity of assumptions}
Eq \eqref{eq:relative} assumes that the calibration data are insensitive to changes in the transmit field across positions such that $\kappa_{k,\text{ref}} = s_1/s_2$. Here we test the validity of this assumption via simulation. Under the small flip angle approximation and allowing for position-specific transmit and receive fields, the calibration images can be written as:
\begin{equation}
    x_k = \frac{s_k R_1 f_{\mathrm{T}_k} \alpha_c \mathrm{TR}_c}{\frac{f_{\mathrm{T}_k}^2 \alpha_c^2}{2} + R_1\mathrm{TR}_c}
    = s_k r
\end{equation}
Here $k$ indexes the repetition of the acquisition, \emph{i.e.} the calibration data for each high resolution VFA acquisition, and $c$ denotes the calibration-specific sequence settings. Considering the ratio method for simplicity, $\kappa_{1,2}$ can then be written more fully as:
\begin{equation}
    \kappa_{1,2} = \Delta_{1,2} \frac{f_{\mathrm{T}_1}f_{\mathrm{T}_2}^2\alpha_c^2 + f_{\mathrm{T}_1}2\mathrm{TR}_c R_1}{f_{\mathrm{T}_2}f_{\mathrm{T}_1}^2\alpha_c^2 + f_{\mathrm{T}_2}2\mathrm{TR}_c R_1}
    = \Delta_{1,2} \frac{\partial \kappa}{\partial\Delta_{1,2}}
\end{equation}
For the ratio of the calibration images, $\kappa_{1,2}$, to equal the relative sensitivity, $\Delta_{1,2}$, we require:
\begin{equation}
    \frac{\partial \kappa}{\partial\Delta_{1,2}}
    = \frac{f_{\mathrm{T}_1}f_{\mathrm{T}_2}^2\alpha_c^2 + f_{\mathrm{T}_1}2\mathrm{TR}_c R_1}{f_{\mathrm{T}_2}f_{\mathrm{T}_1}^2\alpha_c^2 + f_{\mathrm{T}_2}2\mathrm{TR}_c R_1}
    = 1
\end{equation}
We note that the Ernst angle is $\alpha_E = 2\mathrm{TR}_c R_1$, and rearrange to give:
\begin{equation}
    \frac{\alpha_E^2}{\alpha_c^2}\left(f_{\mathrm{T}_2} - f_{\mathrm{T}_1}\right)
    =
    f_{\mathrm{T}_1}f_{\mathrm{T}_2}\left(f_{\mathrm{T}_2} - f_{\mathrm{T}_1}\right)
\end{equation}
This condition is met when $f_{\mathrm{T}_1} = f_{\mathrm{T}_2}$, \emph{i.e.} there is no change in transmit field, or when $f_{\mathrm{T}_1}f_{\mathrm{T}_2}=\alpha_E^2/\alpha_c^2$. These conditions are highlighted in Fig. \ref{fig:kappa_sens} which shows $\frac{\partial \kappa}{\partial\Delta_{1,2}}$ as a function of $f_{\mathrm{T}_j}$ for $R_1 = 0.84s^{-1}$.  Deviation of $\frac{\partial \kappa}{\partial\Delta_{1,2}}$ from 1 is within 3\% for a broad range of values centred on both acquisitions being at the Ernst angle.}

\revised{\textbf{Theoretical error}
Numerically, errors in the $R_1$ estimates were computed for $f_{\mathrm{T}_1} \in [0.5,1.5]$,  $R_1 \in [0.5,1.4]s^{-1}$ and the empirically observed range of relative transmit and receive fields. The median proportion of error arising from transmit or receive field changes was computed over this 4D parameter space.}

\revised{\subsection{Code availability:} The code used to fit the generative model is available at \url{https://github.com/balbasty/multi-bias}. A modified version of the hMRI toolbox that integrates this approach and enables $B_1^+$ correction on a per-contrast basis is available at \url{https://github.com/balbasty/hMRI-toolbox}. The ratio approach can be performed natively with the hMRI toolbox: \url{https://github.com/hMRI-group/hMRI-toolbox}. The source code to reproduce the simulation figures is available at: \url{https://github.com/fil-physics/Publication-Code}.}


%% file: 06_results.tex
\section{Results}

\begin{figure*}[t]
    \centering
    \includegraphics[width=\textwidth]{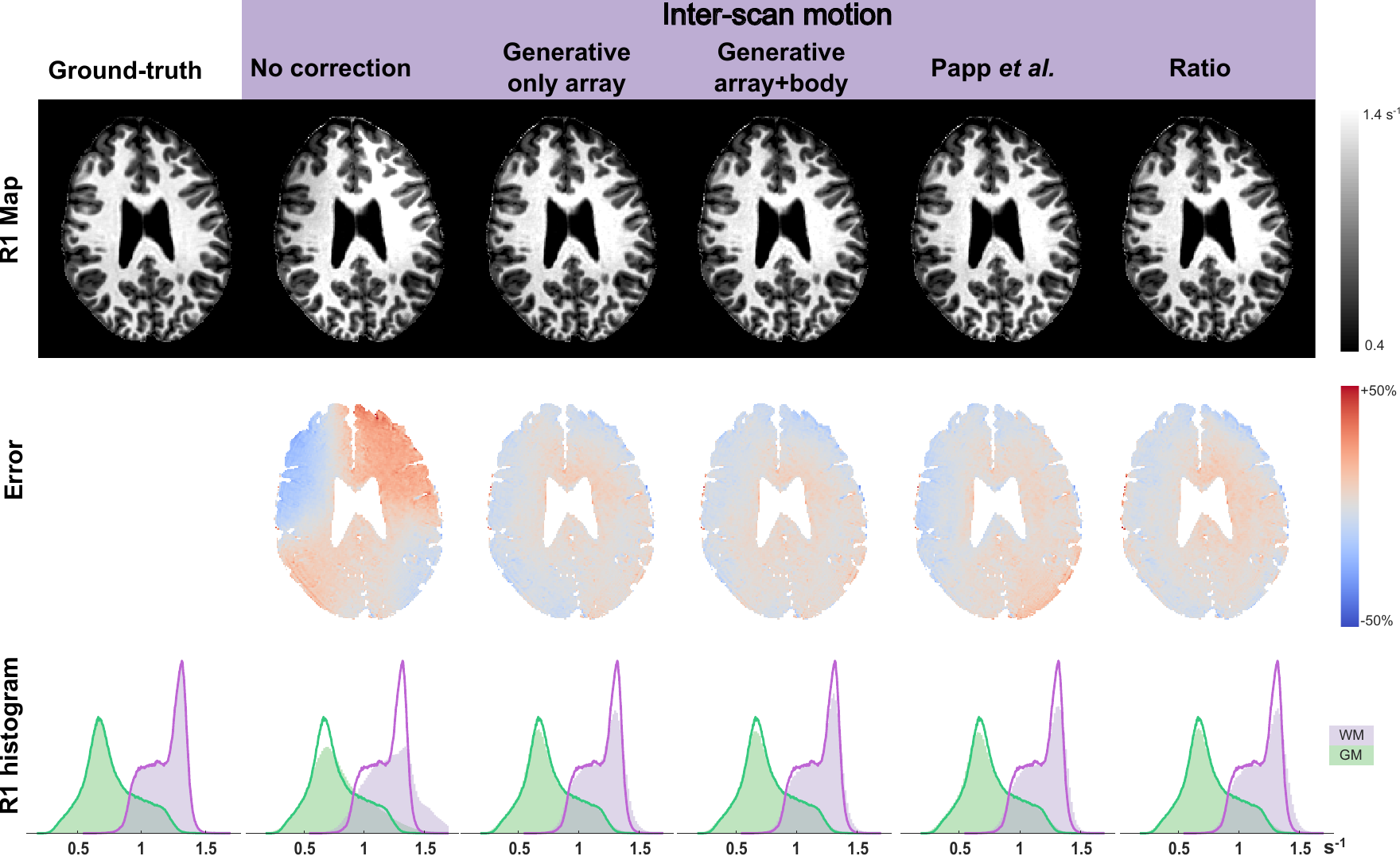}
    \caption{Results at 3T. The first row shows example $R_1$ maps constructed with each method. The second row shows (normalised) error maps with respect to the ground-truth map $\hat{R}_1$. The third row shows histograms (filled area) of $R_1$ within the GM (green) and WM (purple), and their log-Normal fit (solid line); these histograms display probability distributions and therefore integrate to 1. \revised{"Generative / only array" used only the array coil images in the generative modelling framework, whereas "Generative / array+body" incorporated both the array and body coil images using coil-specific regularisation for the smoothness of the sensitivity modulation.}}
    \label{fig:maps3T}
\end{figure*}

\begin{figure*}[p]
    \centering
    \includegraphics[width=\textwidth]{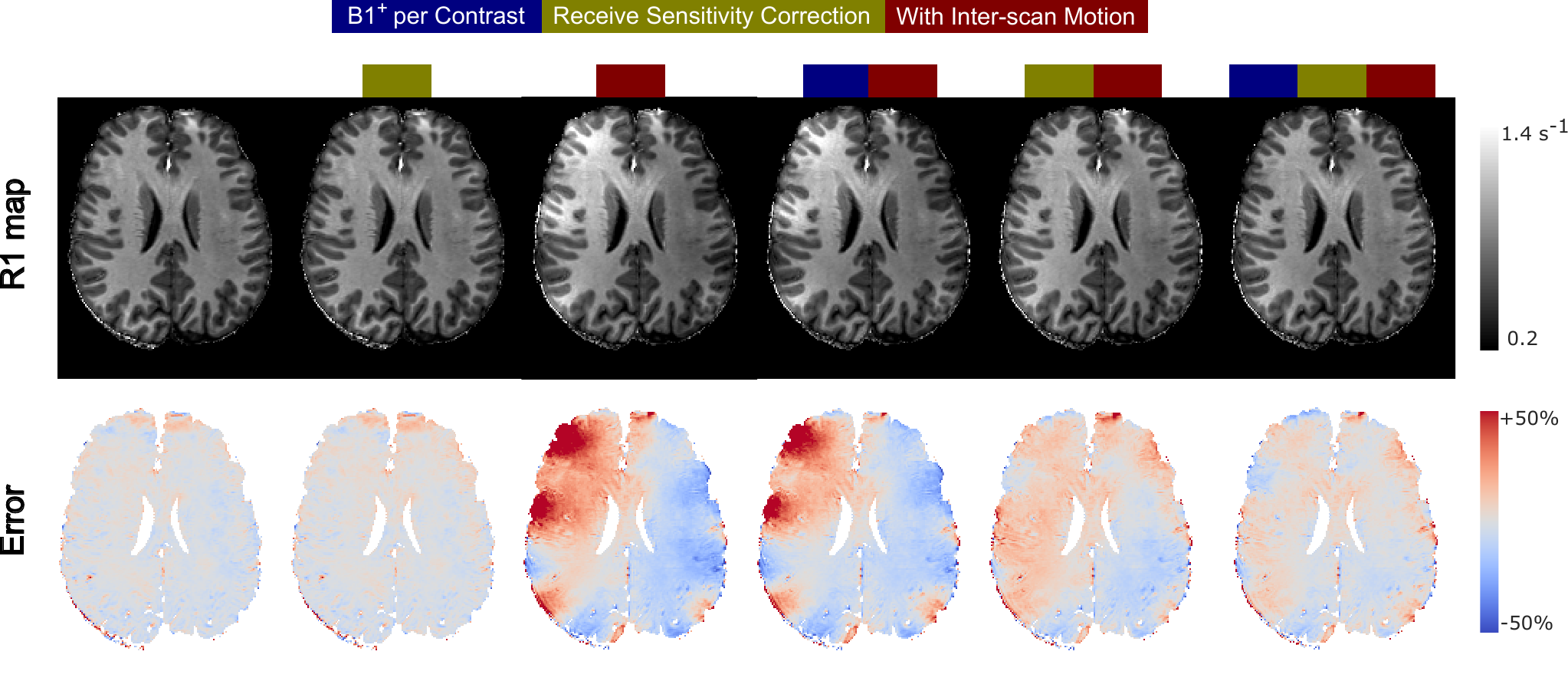}
    \caption{Combining \revised{receive sensitivity} and $B_1^+$ correction at 7T, for participant 3 (last row in figure 3). The first row shows an example $R_1$ map without and with inter-scan motion, before and after net receive sensitivity correction, and employing a separate $B_1^+$ map for each contrast in the case of inter-scan motion. The second row shows (normalised) error maps with respect to the ground-truth map $\hat{R}_1$. In this example, inter-scan motion correction was performed with the generative modelling approach.}
    \label{fig:maps7T_multi}
\end{figure*}

Exemplar images, relative sensitivities, and results from the generative modelling are shown in figure \ref{fig:sens}. The $R_1$ and error maps obtained at 3T and 7T are shown in figures \ref{fig:maps3T} and \ref{fig:maps7T} respectively. The means and standard deviations of the MAE are reported in Table \ref{tab:mae}. The differential impact of correcting for transmit and receive field effects is illustrated in figure \ref{fig:maps7T_multi}. This shows $R_1$ and error maps without motion, and with motion having implemented (i) no correction, (ii) correction only for receive field effects, (iii) only for transmit field effects, or (iv) for both effects in combination.

\subsection{3T Validation}
\revised{The net motion is summarised as the root-sum-of-squares, across the three orthogonal axes, of the translations or rotations independently.  The net translational and rotational motion in the "no motion" condition was 0.8 mm and 0.3 degrees.  These were increased to 1.2 mm and 18.1 degrees in the inter-scan motion case.} When the T1w and PDw volumes were acquired in the same position, the MAE captured the test-retest variability, which was approximately 3\% at 3T and 4-5\% at 7T. In the absence of overt motion, correcting for the differential sensitivity modulation did not substantially change the MAE. In the presence of overt motion, the MAE rose to 10\%.  It was reduced to 4.7\% by the method of Papp \emph{et al.} and to less than 4.4\% by the proposed correction schemes, with or without incorporating the body coil in the generative modelling approach.  The histograms in figure \ref{fig:maps3T} confirm that the method did not introduce any bias to the $R_1$ estimates.

\subsection{Extension to 7T}
\begin{figure*}[p]
    \centering
    \includegraphics[width=\textwidth]{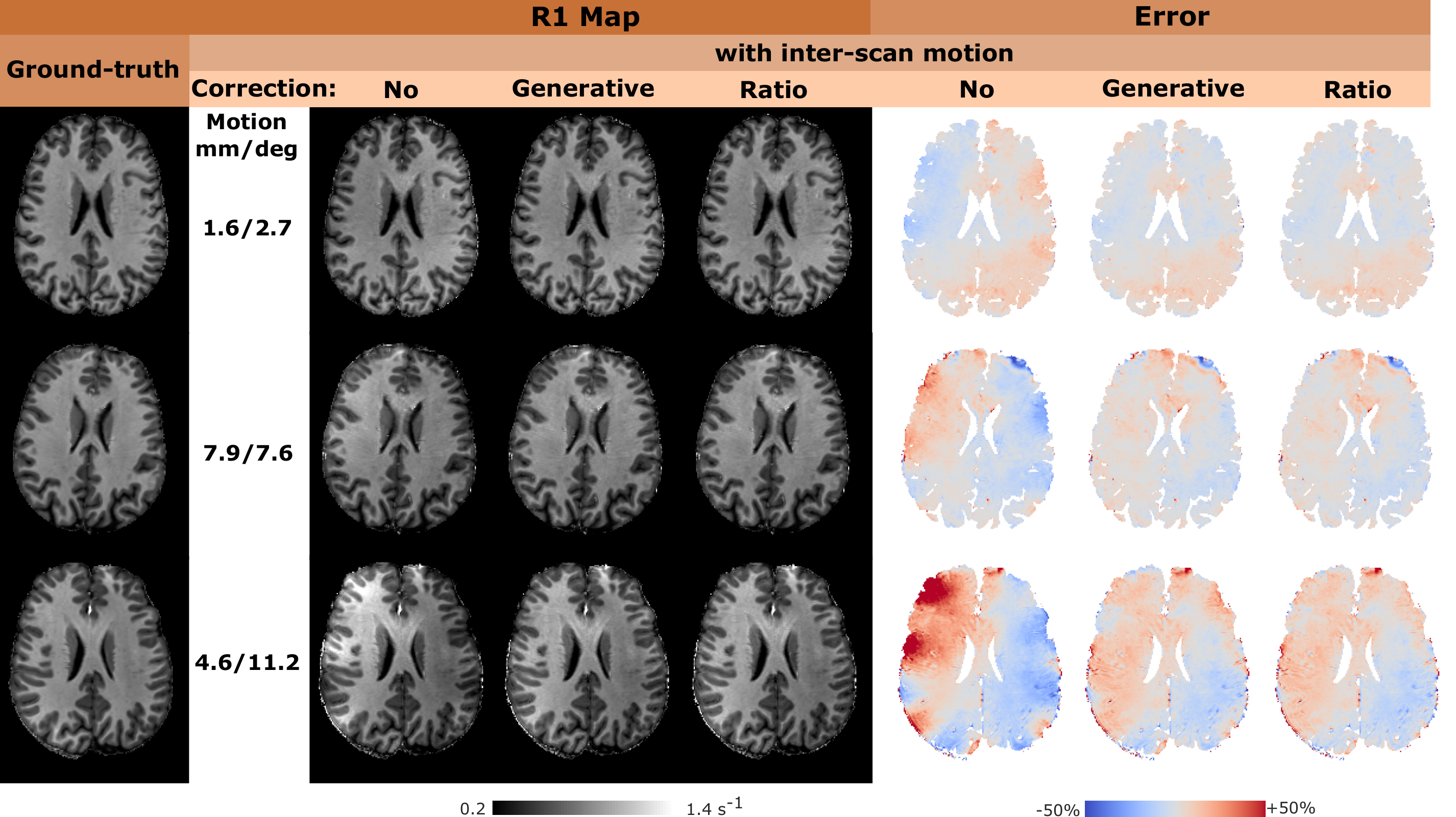}
    \caption{Results at 7T. The first column shows the ground-truth map $\hat{R}_1$ for the three participants sorted based on the magnitude of inter-scan motion. Uncorrected and corrected $R_1$ maps are shown in the middle with the corresponding (normalised) error maps with respect to the ground-truth on the right. Correction is only applied for net \revised{receive sensitivity modulation and not for transmit field effects}. Rows 1 to 3 of the figure correspond to datasets 1, 2 and 3 as reported in Table \ref{tab:mae}. \revised{The net motion is summarised as the root-sum-of-squares, across the three orthogonal axes, of the translations or rotations independently.  In the absence of overt motion, the average displacements between the VFA scans, across the group, were 1 mm and 0.6 degrees for translations and rotations respectively, with a maximum translation of 1.6 mm and a maximum rotation of 1.0 degree.}}
    \label{fig:maps7T}
\end{figure*}

At 7T, the range of motion varied across participants. \revised{The net translational and rotational motion in the "no motion" conditions did not exceed 1.6 mm and 1.0 degree respectively.  In the inter-scan motion cases, the net translation ranged from 1.6 to 7.9 mm, while the net rotation ranged from 2.7 to 11.2 degrees. Rotational motion led to more apparent artefacts.}  The overall amplitude of motion dictated the increase in MAE, which reached a maximum of 13.4\% under the tested conditions (c.f. motion summaries in figure \ref{fig:maps7T} and MAE in table \ref{tab:mae}).  The proposed correction scheme reduced the MAE (5-8\%), though not to the level of no overt motion. 

The variability of the transmit field, $B_1^+$, across head positions was found to be much higher than at 3T. Incorporating position-specific $B_1^+$ maps reduced the MAE (5-12\%) even without correcting for the differential receive sensitivity modulation.  

The greatest reductions in MAE were obtained by correcting for position-specific transmit \emph{and} receive fields, reaching 4-7\%, converging on the level obtained in the absence of overt motion (i.e. 4-5\%). 

\subsection{Comparison of Methods}
Overall, the ratio and generative modelling approaches to \revised{correcting the effects of differential relative sensitivities in $R_1$ maps performed similarly.  The MAE was marginally lower for the generative modelling approach at 3T (0.1\%) and for the ratio approach at 7T (0.5\%).  However, these differences were small relative to the variability across cases (Table \ref{tab:mae})}.

\begin{table*}[t]
\footnotesize  
\caption{MAE (mean $\pm$ s.d. across repeats, in \%) with respect to the average reference $\hat{R}_1$.}
\begin{widetable}{\textwidth}{rcc | cccccc}
    \toprule
    & \textbf{Dataset} & \textbf{Motion} & \textbf{No correction} & \textbf{Ratio} & \textbf{Generative} & \textbf{ \begin{tabular}[c]{@{}c@{}}Generative\\(array + body)\end{tabular}} & \textbf{Papp \emph{et al.}} & \\
    \midrule
    \textbf{3T} & \#1 & No & 3.0 $\pm$ 0.1 & 3.0 $\pm$ 0.1 & 3.0 $\pm$ 0.1 & 3.1 $\pm$ 0.2 & 3.7 $\pm$ 0.2 & \\
    & & Yes & 10.1 $\pm$ 0.8 & 4.4 $\pm$ 0.4 & 4.3 $\pm$ 0.3 & 4.3 $\pm$ 0.1 & 4.7 $\pm$ 0.3 & \\
    \midrule
    & \textbf{Dataset} & \textbf{Motion} & \textbf{No correction} & \textbf{Ratio} & \textbf{Generative} & \textbf{$B_1^+$ per contrast} & \textbf{\begin{tabular}[c]{@{}c@{}}Ratio \&\\ $B_1^+$ per contrast\end{tabular}} & \textbf{\begin{tabular}[c]{@{}c@{}}Generative \& \\ $B_1^+$ per contrast\end{tabular}}\\
    \midrule
    & \#1 & No & 4.0 $\pm$ 0.3 & 4.1 $\pm$ 0.4 & 4.1 $\pm$ 0.3  & 4.1 $\pm$ 0.2  & 4.1 $\pm$ 0.4 & 4.1 $\pm$ 0.3\\
    &     & Yes & 5.8 $\pm$ 0.7 & 4.8 $\pm$ 0.3 & 4.9 $\pm$ 0.4 & 4.9 $\pm$ 0.3 & 3.8 $\pm$ 0.3 & 3.9 $\pm$ 0.3\\
    \textbf{7T} & \#2 & No & 4.8 $\pm$ 0.2 & 4.9 $\pm$ 0.3 & 4.9 $\pm$ 0.3 & 4.8 $\pm$ 0.2 & 5.0 $\pm$ 0.2 & \revised{4.9 $\pm$ 0.2} \\\
    &     & Yes & 8.4 $\pm$ 0.4 & 6.1 $\pm$ 0.3 & 6.2 $\pm$ 0.3 & 8.0 $\pm$ 0.4 & 5.3 $\pm$ 0.1 & 5.5 $\pm$ 0.1 \\
    & \#3 & No & 5.1 $\pm$0.2 & 5.0 $\pm$ 0.2 & 5.0 $\pm$0.2 & 5.0 $\pm$0.2 & \revised{5.1} $\pm$0.3 & \revised{5.1} $\pm$0.2 \\\
    &     & Yes & 13.4 $\pm$ 1.9 & 7.9 $\pm$ 1.1 & 8.3 $\pm$ 1.2 & 11.8 $\pm$ 0.9 & 6.3 $\pm$ 0.3 & 6.8 $\pm$ 0.3  \\
    \bottomrule
\end{widetable}
\label{tab:mae}
\end{table*}

\begin{figure*}
    \centering
    \includegraphics[width=\textwidth]{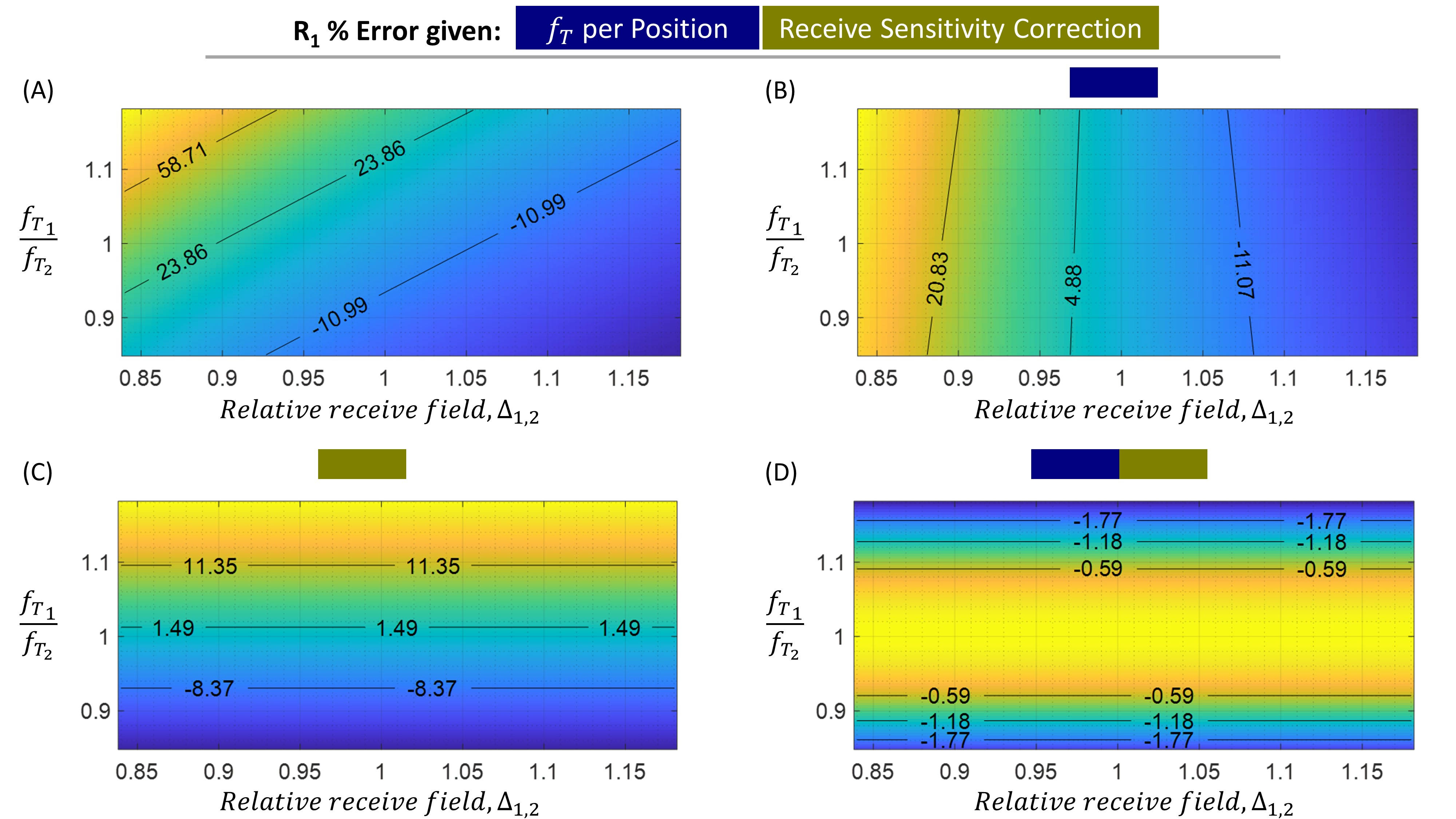}
    \caption{\revised{$R_1$ error (in percentage of the true $\hat{R}_1=0.84s^{-1}$) as a function of the relative transmit field $f_{\text{T}_1}/f_{\text{T}_2}$ and relative receive sensitivity $\Delta_{1,2}$ between two head positions. The four panels show this error with different degrees of correction: (A) none, (B) correction for position-specific transmit field, (C) correction for position specific receive sensitivity, (D) both corrections. Note that minimal error remains even with both corrections, as position-depend transmit effects lead to inaccuracies in the estimation of $\Delta_{1,2}$ by $\kappa_{1,2}$.}}
    \label{fig:numerical_error}
\end{figure*}

\revised{%
\subsection{Numerical $R_1$ Error}
At 3T the relative transmit efficiency ranged from 0.97 to 1.04, whereas the relative receive field (measured via $\kappa_{1,2}$) ranged from 0.84 to 1.18. At 7T the relative transmit efficiency ranged from 0.85 to 1.18 under comparable motion conditions.  These ranges were used in the simulations which revealed that without correction, inter-scan motion caused error as high as 130\%. Over the 4D parameter space investigated, a median of 29\% of the error was caused by transmit field effects and 71\% by receive field effects. Fig. \ref{fig:numerical_error} shows a plane of this error as the relative transmit and receive fields change. Position-specific $f_{\text{T}}$ offers only partial correction (Fig. \ref{fig:numerical_error} B). Larger error reduction arises from receive field correction (Fig. \ref{fig:numerical_error} C). Combining both (Fig. \ref{fig:numerical_error} D) shows receive field effects are removed but transmit sensitivity remains (when $\partial\kappa/\partial\Delta_{1,2} \neq 1$).}

%% file: 07_discussion.tex
\section{Discussion}

We have introduced methods for correcting inter-scan motion artefacts in quantitative MRI that do not rely on the availability or spatial homogeneity of a body coil.  The approaches are based on estimating the \emph{relative} sensitivity modulation across positions, and successfully reduced error in $R_1$ maps at both 3T and 7T.

At 3T, the proposed approaches outperformed a previously established correction method \citep{papp2016correction}. This can be attributed to the fact that the method of Papp \emph{et al.} assumes that the reference modulation of the body coil is independent of position, whereas the proposed methods do not. Instead they specifically account for the \emph{relative} sensitivity across positions thereby restoring consistent modulations.

In the motion conditions tested here, both proposed approaches (ratio with Gaussian smoothing, or generative modelling) produced comparable improvements in $R_1$ reproducibility in the presence of inter-scan motion. Equally importantly, when there was no overt motion neither method decreased reproducibility, which was at a level in keeping with previous reports for similar resolution MPM data \citep{leutritz2020multiparameter, weiskopf2013quantitative}.

\revised{The ratio method benefits from its simplicity, but may be vulnerable to low SNR given that it defines one calibration image as the reference (denominator in equation \eqref{eq:relative}). The alternative generative modelling approach has the benefit of inherently adapts to variable SNR} by estimating the position-specific net sensitivity modulation relative to a common image, which is their barycentre mean. This common image dictates the final modulation of \emph{all} the corrected volumes. The generative model can also easily incorporate any additional data, e.g. body coil images as done at 3T, which flattens the final modulation. Furthermore, rigid registration could be interleaved with model fitting \citep{ashburner2013symmetric} to reach a better global optimum. Finally, the generative model could naturally be integrated with any fitting approach that defines a joint probability over all acquired data, such as \citet{balbastre2021model} in the context of MPM.

The impact of movement on the effective transmit field has previously been investigated in the context of specific absorption rate management \citep{kopanoglu2020specific,le2017probabilistic,wolf2013sar,shajan201416}. An important additional finding of the present work is the impact this can have on $R_1$ estimates at 7T, which was negligible at 3T as demonstrated previously \citep{papp2016correction}.  

\revised{
\subsection{Limitations}
These methods are specifically designed for the correction of inter-scan motion and therefore cannot address intra-scan motion, which may be more likely to occur coincidentally with inter-scan motion, e.g. with uncompliant participants. Although the dominant source of error in $R_1$ was related to receive field effects, the MAE was further reduced by additionally accounting for the positional-dependence of the transmit field. However, acquiring a $B_1^+$ map at multiple positions comes at a cost of increased scan time and inevitably leads to a greater temporal separation between the calibration data and those volumes it is used to correct. Issues such as this, coupled with other uncorrected effects, e.g. position-dependent $B_0$ effects (no reshimming was performed during the experiments), may underlie the fact that the corrections implemented do not reduce the MAE quite to the level of no motion. This finding recapitulates those of Papp et al., though the discrepancy is lower in this work, which is likely because the assumption of a flat body coil receive sensitivity is no longer made.  However, even with the combined receive and transmit field corrections, the MAE is never reduced to the level of no motion.  This is in line with the simulations, which show that position-dependent transmit field effects remain in the calibration data and propagate into the $R_1$ estimates.}

\revised{An additional limitation of the generative model is its reliance on a Gaussian noise assumption, which is violated in the background (but not in the tissue, given the high SNR of the calibration scans). Although we did not find this violation to hamper sensitivity estimation in the present study, the model could nonetheless be modified to incorporate a Rice or noncentral Chi likelihood  \citep{varadarajan2015majorize}.}


%% file: 08_conclusions.tex
\section{Conclusions}
    Inter-scan motion causes serially acquired weighted volumes to be differentially modulated by position-specific coil sensitivities leading to substantial errors when they are combined to compute quantitative metrics. We have demonstrated the efficacy of two methods at reducing these artefacts in the context of $R_1$ mapping. The proposed methods do not require a body coil making them ideally suited for use at 7T, and can be extended to the computation of other quantitative metrics, such as magnetisation transfer saturation \cite{helms2008high}, that similarly assume constant modulation across multiple weighted acquisitions. \revised{Given that the application of this correction does not degrade reproducibility in the no inter-scan motion condition we would recommend it always be used.}


%% file: 10_appendix.tex
\appendix 
\section{Generative Model of Sensitivity-Modulated Images}
The proposed generative model is similar to that of \citet{ashburner2013symmetric} (section 2.3), without non-linear deformations. The magnitude images of the calibration dataset, $\left\{\vec{x}_k \in\mathbb{R}_+^N\right\}_{k=1}^K$, can be written as the voxel-wise product of a mean image $\vec{r}$ and the net sensitivity field $\left\{\vec{s}_k\in\mathbb{R}_+^N\right\}_{k=1}^K$ plus additive noise, approximated as Gaussian with variance $\sigma_k^2$, which is assumed to be uncorrelated across $\vec{x}_k$. The net sensitivity fields can be written as diagonal matrices $\vec{S}_k = \operatorname{diag}\left(\vec{s}_k\right)$, such that the corresponding conditional probability is:
\begin{equation}
    p\left(\vec{x}_k \mid \vec{s}_k, \vec{r}\right) 
    =
    \mathcal{N}\left(\vec{x}_k ~\middle|~ \vec{S}_k\vec{r}, \sigma_k^2\vec{I}\right) ~.
    \label{eq:likelihood}
\end{equation}
If the images have been co-registered but not resliced, a mean space can be defined by computing the barycentre of all aligned orientation matrices \citep{ashburner2013symmetric}. The linear operation of resampling an image from mean to acquired space can be encoded by the matrix $\vec{A}_k\in\mathbb{R}^{N \times N}$ such that the conditional probability becomes:
\begin{equation}
    p\left(\vec{x}_k \mid \vec{s}_k, \vec{r}\right) 
    =
    \mathcal{N}\left(\vec{x}_k ~\middle|~ \vec{A}_k\vec{S}_k\vec{r}, \sigma_k^2\vec{I}\right) ~.
    \label{eq:likelihood_proj}
\end{equation}
While this is the approach taken in practice, the following derivation is restricted to the resliced case for clarity.

The magnitude of each sensitivity field is unknown since the intensity scaling depends on many parameters. However, the sensitivities are known to vary smoothly in space, which is captured by a probability distribution that penalises the field's bending energy \citep{ashburner2007fast}, which integrates the squared curvature of the sensitivity field, making it invariant to intensity scaling. The net sensitivities are encoded by their logs, (\emph{i.e.}, $\vec{s}_k = \exp\vec{z}_k$), such that invariance under shifts in log-space implies invariance under scales in exponentiated space, which is also incorporated into the prior. In a discrete setting, computing the bending energy reduces to the quadratic term $\vec{z}\T\vec{L}\vec{z}$; The prior distribution over sensitivities is therefore defined as a Normal distribution over their logs:
\begin{equation}
    p\left(\vec{z}_k\right)
    =
    \mathcal{N}\left(\vec{z}_k ~\middle|~ \vec{0}, \left(\lambda_k\vec{L}\right)^{-1}\right) ~,
    \label{eq:prior}
\end{equation}
where $\lambda_k$ is an image-specific regularisation factor.

The joint model likelihood, or its negative log, is obtained by combining the likelihood in \eqref{eq:likelihood} and the prior in \eqref{eq:prior}: $\mathcal{L} = -\ln p\left(\left\{\vec{x}_k, \vec{z}_k\right\}_{k=1}^K~\middle|~\vec{r}\right)$. This is minimised with respect to $\left\{\vec{z}_k\right\}_{k=1}^K$ and $\vec{r}$. Neglecting terms that do not depend on these variables, yields the objective function:
\begin{equation}
    \mathcal{L} = \sum_{k=1}^K \left\{ 
    \frac{1}{2\sigma_k^2} \left(\vec{x}_k - \vec{S}_k\vec{r}\right)\T\left(\vec{x}_k - \vec{S}_k\vec{r}\right)
    +
    \frac{\lambda_k}{2}\vec{z}_k\T\vec{L}\vec{z}_k
    \right\} + \text{const} ~.
\end{equation}
Differentiating with respect to the mean image, $\vec{r}$, while keeping the sensitivities fixed gives a voxel-wise ($n$) closed-form update:
\begin{equation}
    r_n \leftarrow \frac{\sum_{k=1}^K s_{kn}x_{kn} / \sigma_k^2}{\sum_{k=1}^K s_{kn}^2 / \sigma_k^2} ~.
\end{equation}
The log-sensitivities have no closed-form solution necessitating an iterative method. The objective function is not everywhere convex, but the likelihood term resembles that of a nonlinear least-squares problem, which can be solved using Gauss-Newton optimisation. Gauss-Newton is a modification of Newton-Raphson that uses Fisher's method of scoring, which amounts to replacing the Hessian at any point with its value at the optimum. With $\vec{R} = \operatorname{diag}\left(\vec{r}\right)$ and $\vec{X}_k = \operatorname{diag}\left(\vec{x}_k\right)$, the gradient $\vec{g}_k$ and Hessian $\vec{H}_k$ used in the Newton-Raphson iteration are:
\begin{align}
    \vec{g}_k & {}= \frac{1}{\sigma_k^2}\vec{S}_k\vec{R}\left(\vec{R}\vec{s}_k - \vec{x}_k\right) + \lambda_k\vec{L}\vec{z}_k ~, \\
    \vec{H}_k & {}=
    \frac{1}{\sigma_k^2}\vec{S}_k\vec{R}^2\vec{S}_k
    + \underbrace{\frac{1}{\sigma_k^2}\vec{S}_k\vec{R}\left(\vec{R}\vec{S}_k - \vec{X}_k\right)}_{\text{Fisher's scoring}~\Rightarrow~\vec{0}}
    + \lambda_k\vec{L} ~.
\end{align}
This ensures that the Hessian used in the Newton-Raphson iteration is positive definite, but does not ensure that the iteration monotonically improves the objective function. We therefore replace the Gauss-Newton Hessian with the more robust preconditioner:
\begin{align}
    \vec{P}_k & {}=
    \frac{1}{\sigma_k^2}\vec{S}_k\vec{R}^2\vec{S}_k
    + \frac{1}{\sigma_k^2}\vec{S}_k\vec{R}\left|\vec{R}\vec{S}_k - \vec{X}_k\right|
    + \lambda_k\vec{L} ~,\\
    \vec{z}_k &{}\leftarrow \vec{z}_k - \vec{P}_k^{-1}\vec{g}_k~,
    \label{eq:update}
\end{align}
which has been shown to yield monotonic convergence \citep{balbastre2021model}. The inversion in equation \eqref{eq:update} is performed with a full multi-grid solver that leverages the sparsity and structure of the preconditioner \citep{ashburner2007fast}.

Finally, a global scaling field $\bar{\vec{s}} = \exp\bar{\vec{z}}$, applied to both the sensitivities ($s_{kn} \leftarrow s_{kn}/\bar{s}_{n}$) and mean image ($r_n \leftarrow \bar{s}_{n}r_n$), ensures that the product $s_{kn}r_n$ is unchanged. Keeping only terms of the objective function that depend on this scaling field gives:
\begin{equation}
    \mathcal{L} \cequal \sum_{k=1}^K \frac{\lambda_k}{2} \left(\vec{z}_k - \bar{\vec{z}}\right)\T\vec{L}\left(\vec{z}_k - \bar{\vec{z}}\right) ~.
\end{equation}
By differentiating, the optimal scaling field is:
\begin{equation}
    \bar{\vec{z}} = \frac{\sum_{k=1}^K \lambda_k \vec{z}_k}{\sum_{k=1}^K \lambda_k} ~.
\end{equation}
Therefore, at the optimum, the (weighted) mean log-sensitivity field must be zero, and the mean image is a barycentre of the calibration images. To accelerate convergence, this condition is enforced after each global iteration.
